\documentclass[graybox]{svmult}
\usepackage{type1cm}        
\usepackage{makeidx}         
\usepackage{graphicx}        
\usepackage{multicol}        
\usepackage[bottom]{footmisc}

\usepackage{newtxtext}       %
\usepackage{newtxmath}       

\usepackage{graphicx}
\usepackage{epstopdf}
\usepackage[export]{adjustbox}
\graphicspath{{"fig/"}}
\usepackage{subcaption}

\usepackage{natbib}
\setcitestyle{authoryear,open={(},close={)}}

\usepackage{multicol}
\usepackage{multirow}
\usepackage{booktabs}
\usepackage{caption}
\usepackage{threeparttable}
\usepackage[justification=centering]{caption}

\makeindex             

\usepackage{hyperref}
\begin{document}

\title*{To what extent airborne particulate matters are influenced by ammonia and nitrogen oxides?}
\titlerunning{To what extent PM2.5 is due to NH3 and NOX?}
\author{Alessandro Fassò}
\institute{Alessandro Fassò \at University of Bergamo, Via dei Caniana, 2, Bergamo, Italy, \email{alessandro.fasso@unibg.it}}
%
%
\maketitle

\abstract{
Intensive farming is known to significantly impact air quality, particularly fine particulate matter (PM$_{2.5}$). Understanding in detial their relation is important for scientific reasons and policy making.
\\
Ammonia emissions convey the impact of farming, but are not directly observed. They are computed through emission inventories based on administrative data and provided on a regular spatial grid at daily resolution.
\\
In this paper, we aim to validate \textit{lato sensu} the approach mentioned above by considering ammonia concentrations instead of emissions in the Lombardy Region, Italy. While the former are available only in few monitoring stations around the region, they are direct observations.
\\
Hence, we build a model explaining PM2.5 based on precursors, ammonia (NH3) and nitrogen oxides (NOX), and meteorological variables. 
To do this, we use a seasonal interaction regression model allowing for temporal autocorrelation, correlation between stations, and heteroskedasticity.
\\
It is found that the sensitivity of PM2.5 to NH3 and NOX depends on season, area, and NOX level.
It is recommended that an emission reduction policy should focus on the entire manure cycle and not only on spread practices.
} 

\section{Introduction} \label{Introduction}

Intensive farming is known to impact air quality severely because of ammonia emissions.
In fact, the fertilisation of fields and, more importantly, manure management release ammonia into the atmosphere. 
Here, after a number of chemical reactions, a large part of ammonia is converted into secondary fine particulate matter (PM$_{2.5}$).
The contribution of ammonia emitted by livestock to PM2.5 has been studied from different perspectives in recent decades. 

\cite{HRISTOV2011} considered a chemical perspective in the United States and found that ammonia contribution was between 5 and 20\%. 
Chemical transport model (CTM) simulations have been applied to an ad hoc emission inventory by \cite{Thunis2021} and \cite{Veratti2023}. 
Focusing on the Po Basin, Italy, they considered the interactions between nitrogen oxides (NOX) and ammonia (NH3) and showed that a 25\% reduction of NOx and NH3 results in a reduction (sensitivity) of $\approx$ 10 $\mu g/m^3$ during winter.
\cite{Nenes2020} consider the importance of humidity in determining the sensitivity of PM2.5 to ammonia.

An alternative approach has been developed in the AGRIMONIA project (www.agrimonia.net), which considered the Copernicus NH3 emission data set in its open-access data set \citep{fasso2023agrimonia}.
Along these lines, \cite{agrimonia_IES2023} and \cite{otto2023spatiotemporal} considered the relation between PM2.5 and ammonia emissions using advanced statistical and machine learning models in the Lombardy Region, Italy.
A hybrid machine learning-causal inference approach was used by \cite{SongEtAl_2023} for policy assessment.
Essentially, PM2.5 is used as the response variable of a spatiotemporal regression with predictors covering meteorological variations, which is known to condition pollutant concentration, and including ammonia emissions, which are available on a regular spatial grid with low temporal and spatial resolution.

The issue of the quality of the emission inventory is quite challenging. Machine learning has been used to estimate ammonia emissions \citep{Hempel2020} and/or calibrate existing ammonia emission inventories to available ammonia concentration data \citep{NH3_via_ML_preprint}.
Unfortunately, ammonia concentrations are not extensively monitored in Europe.
A notable exception is Switzerland, where meteorologically adjusted trends in the years 2000-2021 are available for a number of monitoring stations \citep{grange2023NH3inCH}.

In this paper, we aim to validate \textit{lato sensu} the approach of \cite{otto2023spatiotemporal} using field observation of ammonia and nitrogen oxide concentrations. 
Whilst ammonia emission data has an extensive spatial coverage \citep{Agrimonia_dataset}, ammonia concentration observations are available at a limited number of locations in Lombardy.
Hence, to handle this data, a model simpler than in \cite{otto2023spatiotemporal} is used here.

In the sequel, a seasonal regression model with interactions, autocorrelation in time, correlation between stations, and heteroskedastic errors is used.
The interaction and error structures are selected among a large number of alternative models. Due to the computational burden of the large data set, a new simplified estimation algorithm is proposed.

The selected model is used to answer the following scientific questions:
\begin{itemize}
\item[SQ1.] \hspace{3mm} Is PM2.5 sensitive to NH3 and NOX?
\item[SQ2.] \hspace{3mm} Does this relation depend on the Season?
\item[SQ3.] \hspace{3mm} Does the sensitivity to NH3 depend on NOX level?
\item[SQ4.] \hspace{3mm} Is this relation constant around Lombardy?
\end{itemize}

The rest of this article is organised as follows.
Section \ref{sec:data} describes the subset of the AGRIMONIA data set composed of the five stations measuring both PM2.5 and NH3. Moreover, a preliminary analysis addresses some data features that are important for modelling.
In Section \ref{sec:modelling}, the statistical model is introduced and an estimation algorithm is proposed.
In Section \ref{sec:fitting}, an extensive model selection is performed and the results are presented.
Section \ref{sec:discussion} provides a discussion of the results and answers the above scientific questions.
A concluding section closes the paper.

\section{Data}\label{sec:data}
This paper is based on the AGRIMONIA data set \citep{Agrimonia_dataset}, extensively discussed by \cite{fasso2023agrimonia}, which collects harmonised daily data on air quality, meteorology, and agriculture in Lombardy, Italy, for the years 2016-2021.
In particular, we focus on the five stations that simultaneously provide daily data on fine particulate matter (PM2.5), ammonia (NH3), and nitrogen oxides (NOX) with good coverage over the entire period.

The air quality network behind AGRIMONIA was designed by the Lombardy Environment Protection Agency (ARPA Lombardy), with little consideration of agriculture emissions and ammonia pollution. As a result, intensive farming areas are under-sampled; see Figure 3 in \cite{fasso2023agrimonia}. 
Ammonia sampling designs are quite heterogeneous in continental Europe, with some countries - e.g., Switzerland \citep{grange2023NH3inCH} -  implementing extensive long-term monitoring and others doing very little.
The Italian situation is partly a consequence of EU air quality monitoring legislation, which does not require systematic air quality monitoring in connection to livestock farm emissions.
As a result, ammonia concentrations are not provided by the European Environment Agency, and only a few regions voluntarily collect ammonia at very sparse monitoring stations.

\begin{figure}
\centering
\includegraphics[width=1\textwidth]{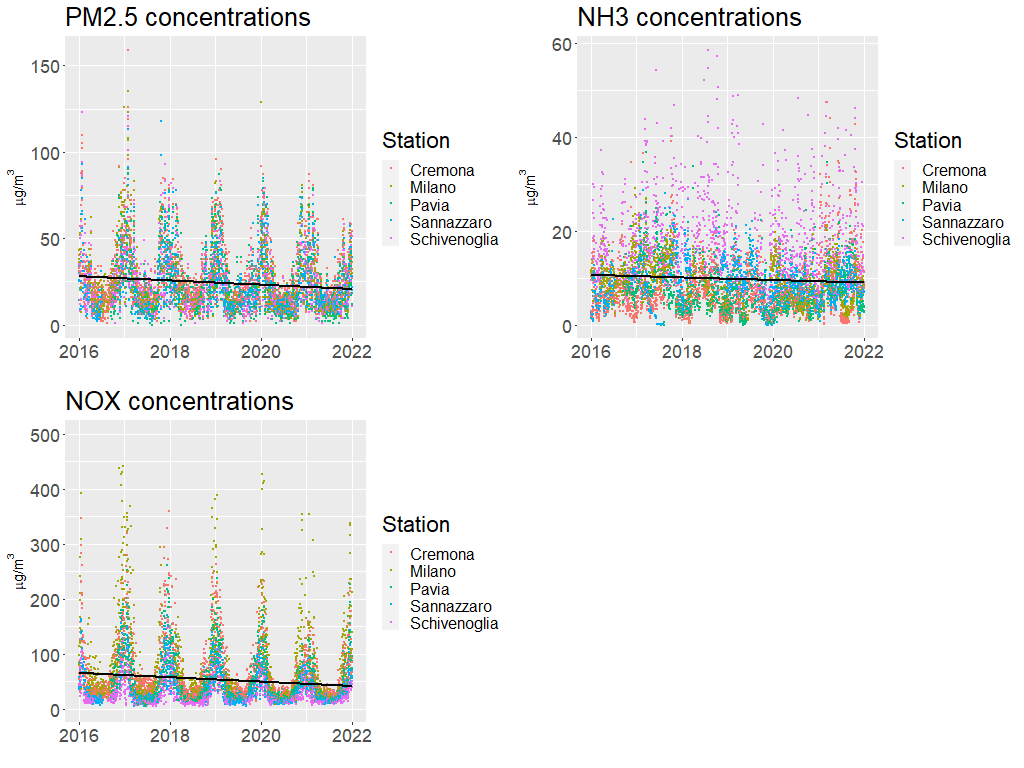} 
\caption{Airborne pollutant daily time series for the considered five stations.}
\label{fig:AQ_timeseries}
\end{figure}

Figure \ref{fig:AQ_timeseries} depicts the air quality data used in this paper, 
and Table \ref{tab:elenco_stazioni} reports some additional details. 
It is seen that that only one station, Schivenoglia, is located in a rural area, with very intensive swine farming activity. Indeed it has an ammonia average of about twice the other ones.

Milano station is in a large metropolitan area, far away from livestock and agriculture activity.
The other four stations are located in relatively small cities surrounded by agricultural activity, South of Milano, in Lower Po Valley where the air circulation is limited.
Only one is a rural background (RB) station, but all are expected to be influenced by agricultural activity.
We will refer to them as Lower Po Valley (LPV) stations in the sequel.

Due to about 30\% of missing days, the six-year data set has $N=7770$ observations, with the numbers per station given in Table \ref{tab:elenco_stazioni}. 
The presence of missing data does not pose a numerical problem for the model used in the next section. 
In fact, with regard to possible biases introduced by an unbalanced missing pattern, we observe that two stations, Moggio (LC) and Bergamo (BG), have been removed because they provided little data and only in recent years. 
The other five stations present a good seasonal balance and a good coverage of the years considered.

\begin{table}[ht]
\centering
\begin{tabular}{lccccccc}
  \hline
 Station location & ID & Province & Type & PM2.5 & NH3 & NOX & $n$ \\ 
  \hline
Cremona, Via Fatebenefratelli & 677 & CR & UB & 26.45 & 7.58 & 62.18 & 1923 \\ 
Milano, Via Pascal, Citta Studi & 705 & MI & UB & 23.15 & 9.11 & 78.98 & 1358 \\ 
Pavia, Via Folperti & 642 & PV & UB & 23.98 & 7.55 & 53.53 & 1547 \\ 
Sannazzaro De Burgondi, Via Traversi & 693 & PV & UI & 22.73 & 8.54 & 39.54 & 1570 \\ 
Schivenoglia, Via Malpasso & 703 & MN & RB & 23.20 & 17.16 & 27.02 & 1372 \\ 
   \hline
\end{tabular}
\caption{ARPA Stations used in this paper. ID is the unique ARPA identifier; Type acronyms UB, UI, and RB denote Urban Background, Urban Industrial, and Rural Background, respectively. PM2.5-NOX columns report the averages, and $n$ is the number of available observations per station.}
\label{tab:elenco_stazioni}
\end{table}

\begin{figure}
\centering
\includegraphics[width=1\textwidth]{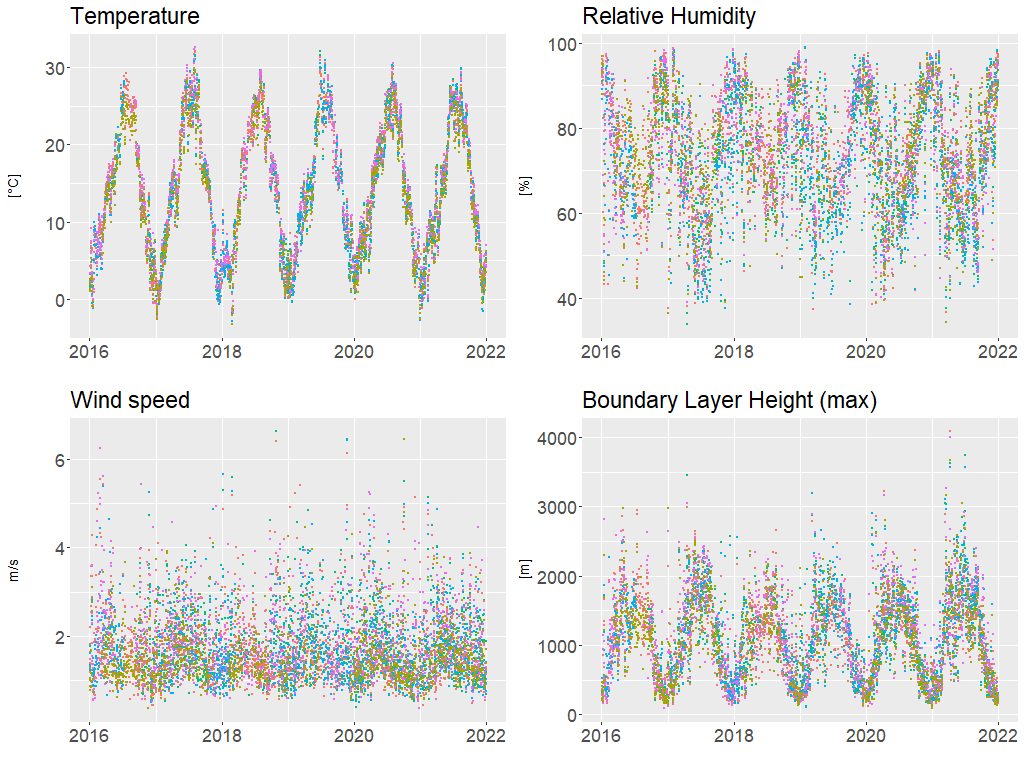} 
\caption{Meteorological daily time series. Station colours are the same as in Figure \ref{fig:AQ_timeseries}.}
\label{fig:WE_timeseries}
\end{figure}

Meteorological predictors used in this paper 
to take into account pollutant variations due to meteorology
are temperature (T), relative humidity (RH), total precipitation, wind speed (WS), and boundary layer height (BLH) used in the subsequent modelling; see Figure \ref{fig:WE_timeseries}.

\subsection{Preliminary analysis}\label{sec:prelim.analysis}
Most of the variables in Figures \ref{fig:AQ_timeseries} and \ref{fig:WE_timeseries} clearly show a strong seasonal pattern for all stations, and the three air quality variables show a slight decreasing trend. In particular, ammonia shows less pronounced seasonality and trend. Table \ref{tab:nh3_by_season} shows that the winter average is generally smaller. In LPV, this is consistent with reduced agriculture activity during this season. In contrast, it is larger at the Milano station due to the contribution of urban traffic to ammonia.

\begin{table}[ht]
\centering
\begin{tabular}{lrrrr}
  \hline
 & Spring & Summer & Fall & Winter \\ 
  \hline
Cremona & 7.406 & 7.669 & 9.034 & 6.308 \\ 
  Milano & 9.863 & 7.668 & 8.905 & 9.954 \\ 
  Pavia & 7.710 & 9.430 & 7.638 & 5.797 \\ 
  Sannazzaro & 7.567 & 8.762 & 9.783 & 8.389 \\ 
  Schivenoglia & 16.438 & 20.363 & 16.608 & 16.344 \\ 
   \hline
\end{tabular}
\caption{NH3 average by season and station.}
\label{tab:nh3_by_season}
\end{table}

\begin{figure}
\begin{tabular}{cc}
\includegraphics[width=.55\textwidth]{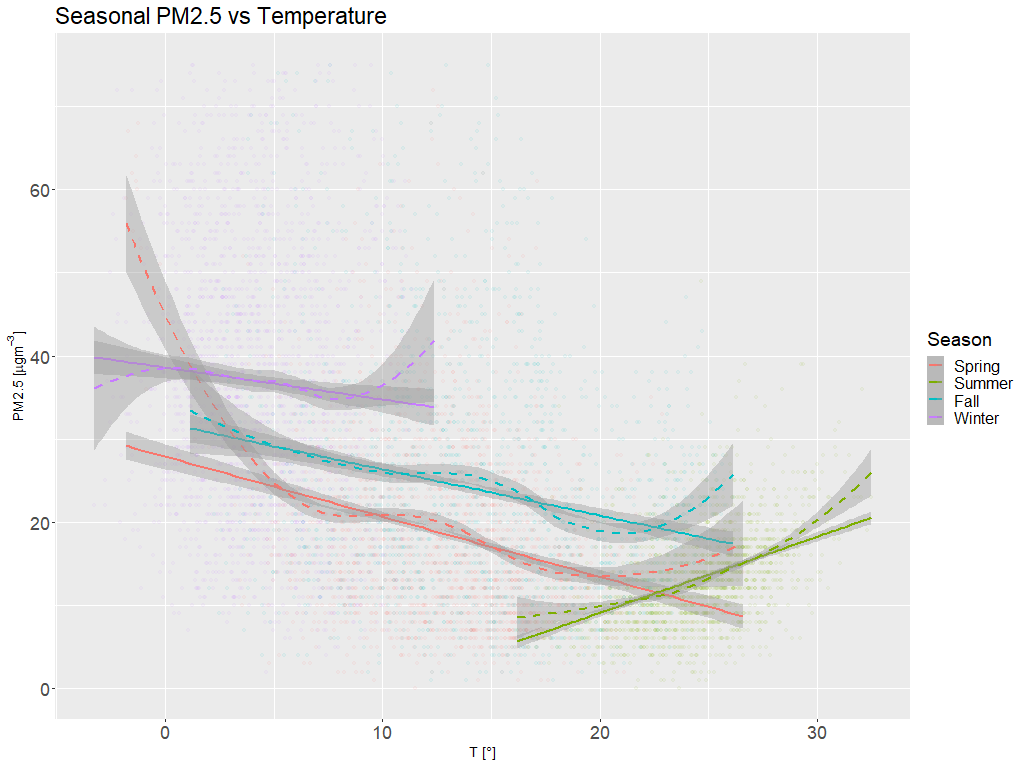}  &
\includegraphics[width=.55\textwidth]{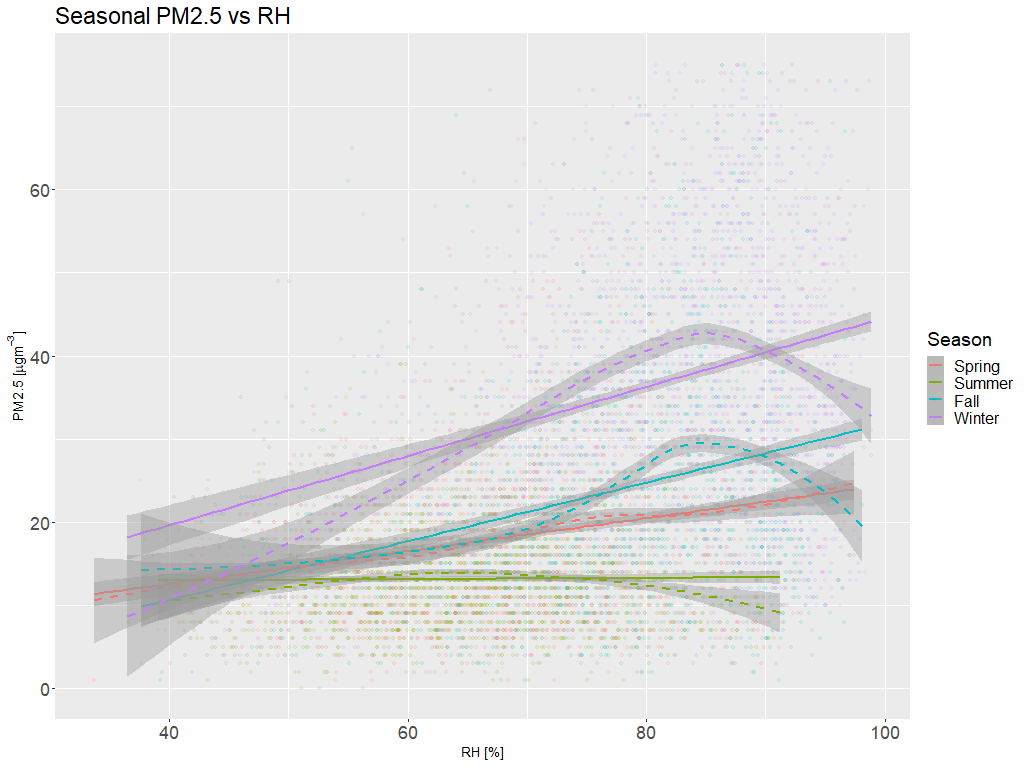}  \\
\includegraphics[width=.55\textwidth]{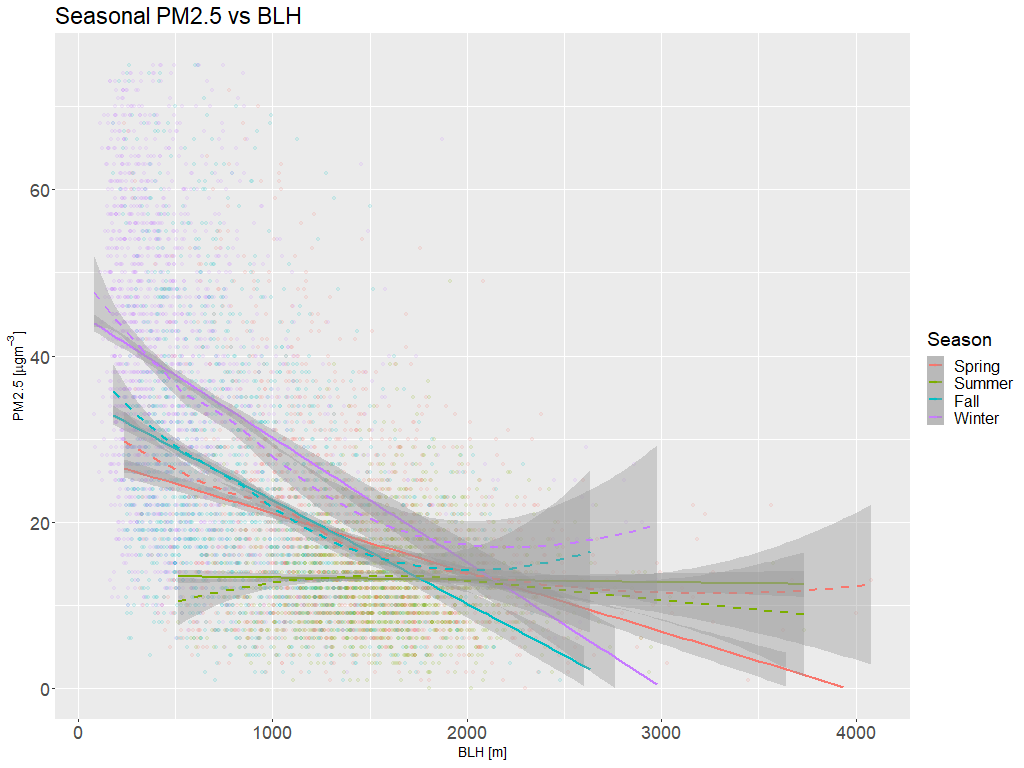}  &
\includegraphics[width=.55\textwidth]{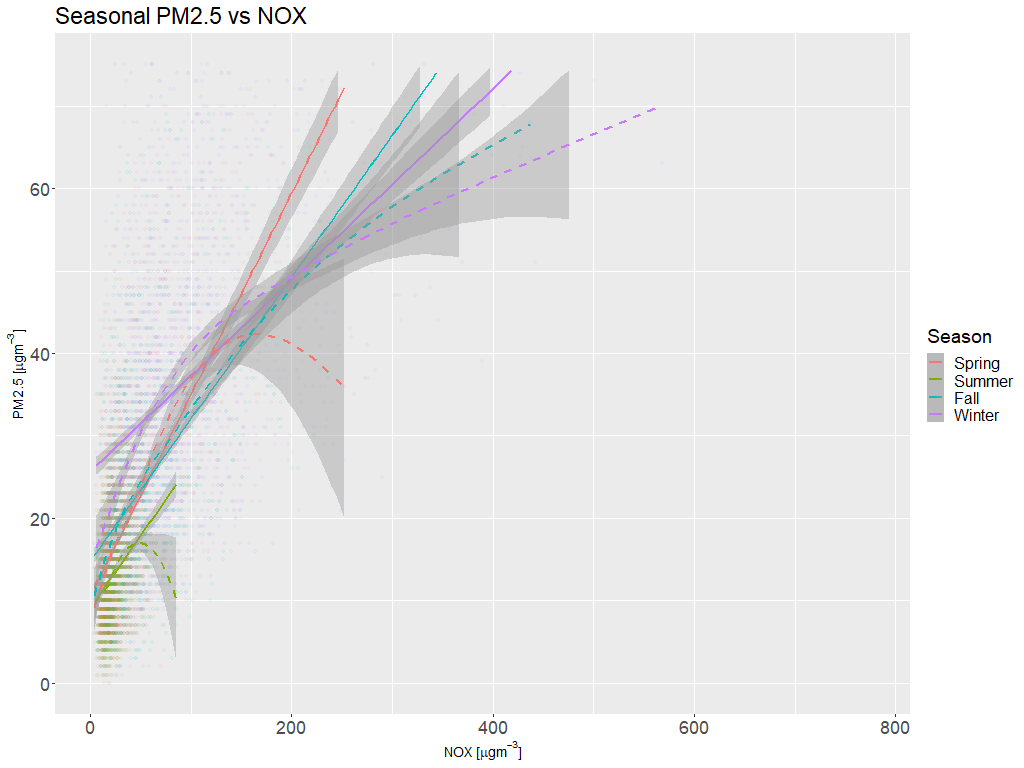}  \\
\includegraphics[width=.55\textwidth]{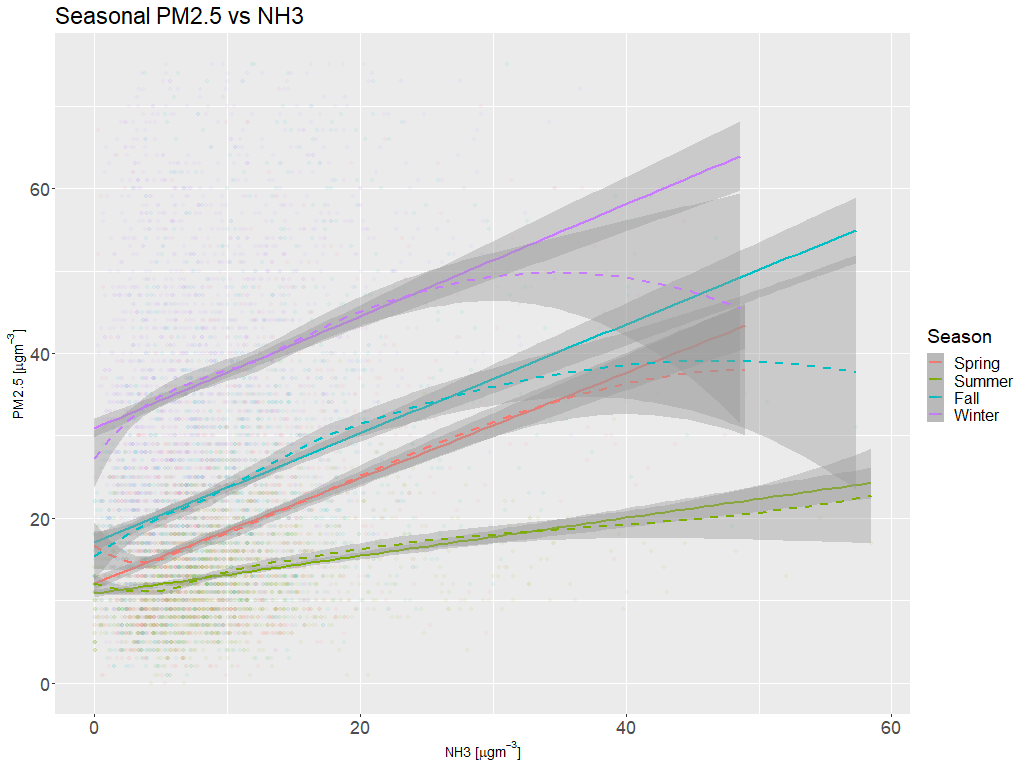}  &
\\
\end{tabular}
\caption{Seasonal plots of PM2.5 versus T, RH, BLH, NOX, and NH3 on zigzag order from top-left to bottom-right. The solid lines are least square fits and the dashed lines are LOESS fits. Shaded areas are 95\% bands.}
\label{fig:SeasonalPM25vsALL}
\end{figure}

One might wonder whether the relation between the response (PM2.5) and the predictors (NH3, NOX, T, RH, WS, and BLH) is linear.
Figure \ref{fig:SeasonalPM25vsALL} depicts linear and LOESS fit by season, and shows that the relationship between PM2.5 and predictors depends on the season for both meteorological variables and pollutants. 
In some cases, it is also non-linear after fixing the season.

In particular, T has a generally negative correlation with PM2.5, but such correlation is positive in summer. 
Moreover, in spring, the slope is steeper at low T than at medium and high T.
The case of RH is known in the literature - e.g., \cite{Vaishali2023}. In the top-right panel of Figure \ref{fig:SeasonalPM25vsALL}, we see a deflection of PM2.5 at high RH.

\begin{figure}
\centering
\includegraphics[width=.9\textwidth]{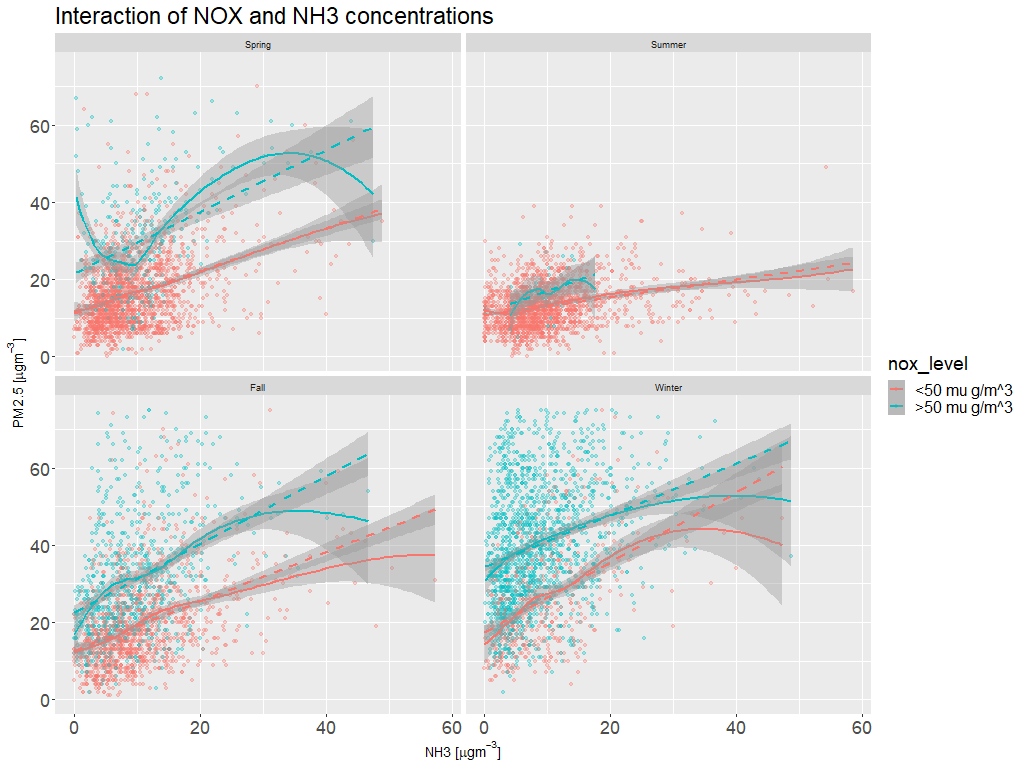} 
\caption{PM2.5 versus NH3 by season and level of NOX. The solid lines are least square fits and the dashed lines are LOESS fits. Shaded areas are 95\% bands. The cyan color refers to data with high NOX (larger than 50 $\mu g m^{-3}$), and the orange refers to low NOX level.}
\label{fig:InteractionNH3-nox4}
\end{figure}

Also, the relation between PM2.5 and NH3 has been recognised to depend on the NOX level - e.g., \cite{Veratti2023}.
Some evidence of this is given in Figure \ref{fig:InteractionNH3-nox4}, where cyan refers to data with high NOX (greater than 50 $\mu g m^{-3}$), and the orange refers to low NOX level. This suggests some similarities to the behaviour of PM2.5 under air stagnation in China \citep{Zhang_airStagnation_2023}.

\section{Statistical modelling}\label{sec:modelling}
In this paper, we use a seasonal linear model with serial correlation, correlation among stations, and heteroskedasticity in the frame of generalised linear models \citep{pinheiroBates2006}.

Let $y_{s,d}$ denote PM2.5 observation at day $d=1,\dots,n_d$ and station $s=1,\dots,n_s$. Then, we consider the local linear model
\begin{equation}\label{eq:Model1}
y_{s,d} = c + \alpha y_{s,d-1} + \beta(z_{s,d})' x_{s,d} + \varepsilon_{s,d}
\end{equation}
where $\alpha$ is the first-order autoregressive coefficient and determines the persistency of $y$. The predictor vector $x$ includes meteorological variables and PM2.5 precursors (NH3 and NOX).
The vector $\beta$ depends on $z$, which may include season, NOX level being above or below 50 $\mu g m^{-3}$, and geographical information.
Contemporaneous errors in different stations, say $\varepsilon_{s,d}$ and $\varepsilon_{s',d}$, are correlated. We denote by $R =(\rho_{s,s'})$ the related $n_s \times n_s$ correlation matrix.

The Equation (\ref{eq:Model1})'s error variance, $\tau^2_{s,d}=Var(\varepsilon_{s,d})$, is allowed to change with the season using the following skedastic model
\begin{equation}\label{eq:var.seasonal}
\tau_{s,d} = \sigma \gamma_\mathcal{S}
\end{equation}
where $\mathcal{S}=1, \dots, 4$ is the season of day $d$, $\sigma$ and $\gamma_\mathcal{S}$ are positive, and $\gamma_1=1$ for identifiability.
Also, we consider a more general skedastic model where the error variance $\tau^2$ depends on the conditional expected PM2.5 level $\hat{y}_{s,d}=E(y_{s,d}|y_{s,d-1},x_{s,d})$.
In this case, we use the general power model
\begin{equation}\label{eq:varPower}
\tau_{s,d} = \sigma(\gamma_\mathcal{S} + \hat{y}_{s,d}^{\delta_\mathcal{S}}).
\end{equation}
An intermediate case between Equations (\ref{eq:var.seasonal}) and (\ref{eq:varPower}) uses a constant power $\delta_\mathcal{S}=\delta$.
In the rest of the paper, with abuse of notation, we will use $\gamma$ for the vector with components $\gamma_1 \cdots, \gamma_4$ or, in case of a non-seasonal model, for the corresponding scalar value. A similar convention will apply to $\delta$.

Equation \ref{eq:Model1} can be interpreted as follows. The daily PM2.5 level depends on the level of the previous day plus a variation which depends on local meteorology and precursors. Of course, both data and model information contents are limited. 
In particular, precise time dynamics and transport may influence PM.5. 
It is therefore appropriate to admit that the error has a spatial structure - i.e., it has a hidden component shared with other stations. 
Moreover, the error size may depend on the season and may be proportional to the expected PM2.5 level.

If we take $z_{s,d}$ to describe the season and NOX level, an interesting special case of Equation (\ref{eq:Model1}) can be rewritten using an R-like Wilkinson notation as follows:
\begin{equation}\label{eq:Model.Wilk}
\texttt{PM2.5} \sim \texttt{1 + Season:(lag(PM2.5)+NH3:nl+NOX+meteo)}
\end{equation}
where \texttt{Season} is a four-level factor, \texttt{nl} is a dummy variable for the high NOX level, and \texttt{meteo} includes the meteorological variables discussed in the previous section.
This model will be the basis for the model selection we perform below.
\subsection{Estimation}
Estimation of models such as those in Equations (\ref{eq:Model1}) and (\ref{eq:Model.Wilk}) can be performed using the maximum likelihood method, thanks to the \texttt{nlme} package available in the Cran Archive (\url{https://cran.r-project.org}).
Unfortunately, the model complexity related to some fifty parameters and $N=7770$ observations makes simultaneous estimation of all parameters infeasible on a recent, high-performance laptop computer.
Hence, the following multi-step algorithm has been devised:\vspace{2mm} \\

\noindent
\textbf{Algorithm 1}\vspace{-6mm} \\
\begin{enumerate}
\item Ordinary Least Square (OLS) estimation of the regression component, say $\hat{\beta}^0$.
\item Empirical estimation of error seasonal variances, say 
$\hat{\gamma}^0$, based on OLS residuals.
\item Maximum likelihood estimation of the inter-station correlation matrix, say $\hat{R}$,
assuming known seasonal variances $\hat{\gamma}^0$.
\item Maximum likelihood estimation of $\beta,\gamma$, and $\delta$,
assuming known correlation matrix $\hat{R}$ and initial values $\hat{\gamma}^0$ for $\gamma$.
\end{enumerate}

\section{Model selection and fitting}\label{sec:fitting}
In this section, we consider model selection regarding both the regression part, described in Equations (\ref{eq:Model1}) and (\ref{eq:Model.Wilk}), and the skedastic component, described in Equations (\ref{eq:var.seasonal}) and (\ref{eq:varPower}).

One component that is always included in the model is the matrix of error correlations between stations. In fact, missing this component may affect the inference validity.
Furthermore, a ``basic'' regression part - denoted by $X$ in the sequel -  always included in the model is as follows: lagged PM2.5, NH3 interacted with NOX level, i.e. the component \texttt{NH3:nl} of Equation (\ref{eq:Model.Wilk}), four meteorological predictors T, RH, WS, and BLH of Section \ref{sec:prelim.analysis} plus a dummy variable identifying rainy days.

Keeping the above $X$ as a common component, we build the ten interaction model variations listed in Table \ref{tab:ten.models} by increasing complexity. From Models 1 to 4, the role of seasonality is considered by adding an additive seasonal component in Model 2, a purely multiplicative seasonal model (symbol \texttt{:}) in Model 3, and a fully multiplicative seasonal model (symbol \texttt{*}) in Model 4. 
Models 5 to 10 take into account the role of the dummy for Milano station and its interaction with seasonality. 
The introduction of the dummy variable for Milano station in Models 5-10 is motivated by the unreported results of Models 1-4. For these models, we found that the residual autocorrelation by station is generally good, except for Milano station, where an important autocorrelation at lag one is observed.

\begin{table}[ht]
\centering
\begin{tabular}{cl}
  \hline
Model ID & Formula \\ 
  \hline
 1 & PM25 $\sim$ X \\ 
 2 & \texttt{PM25} $\sim$ \texttt{X + Season} \\ 
 3 & \texttt{PM25} $\sim$ \texttt{X : Season} \\ 
 4 & \texttt{PM25} $\sim$ \texttt{X * Season} \\ 
 5 & \texttt{PM25} $\sim$ \texttt{X + Milano} \\ 
 6 & \texttt{PM25} $\sim$ \texttt{X + Season + Milano} \\ 
 7 & \texttt{PM25} $\sim$ \texttt{(X : Season) : Milano} \\
 8 & \texttt{PM25} $\sim$ \texttt{X : Season  *  Milano} \\ 
 9 & \texttt{PM25} $\sim$ \texttt{(X * Season): Milano} \\ 
10 & \texttt{PM25} $\sim$ \texttt{(X * Season)* Milano} \\ 
   \hline
\end{tabular}
\caption{Regression models considered. \\
\texttt{X = lag(PM2.5)+NH3:nl+NOX+T+RH+WS+BLH+RainyDay}}
and \texttt{Milano} is a dummy for Milano station.
\label{tab:ten.models}
\end{table}
	
Regarding the skedastic function in Equation \ref{eq:varPower}, we consider four alternatives, moving from more constrained and parsimonious models to the saturated one:
\begin{enumerate}
\item[-] Model 1: The pure seasonal variance given by Equation \ref{eq:var.seasonal}. 
\item[-] Model 2: Non-seasonal power variance given by the Equation \ref{eq:varPower} (omitting the indicator $\mathcal{S}$), with estimated constant $\gamma$ and fixed power $\delta=1$.
\item[-] Model 3: Seasonal power variance given by Equation \ref{eq:varPower} with estimated constants $\gamma_\mathcal{S}$ and fixed power $\delta=1$.
\item[-] Model 4: Seasonal power variance given by Equation \ref{eq:varPower} with estimated constant and power vectors $\gamma$ and $\delta$.
\end{enumerate}

In the sequel, we will identify a model by the two indexes $(i,j)$, with $i=1, \dots, 10$, and $j = 1, \dots, 4$. 
Tables \ref{tab:AIC} and \ref{tab:BIC} report the Akaike Information Criterion (AIC) and the Bayesian Information Criterion (BIC), respectively \citep{AIC,BIC}.\\

\begin{table}[ht]
\centering
\begin{tabular}{rllcccccccc}
& && \multicolumn{8}{c}{Skedastic model} \\
\cline{4-11}
& && \multicolumn{2}{c}{1} 
  & \multicolumn{2}{c}{2} 
  & \multicolumn{2}{c}{3}
  & \multicolumn{2}{c}{4}\\
\multicolumn{2}{c}{Regression model} && AIC & df & AIC & df & AIC & df & AIC & df \\ 
  \hline
 1 & X                   && 51660 & 14 & 50710 & 12 & 50683 & 15 & 50676 & 19 \\ 
 2 & X + Season          && 51596 & 17 & 50664 & 15 & 50631 & 18 & 50608 & 22 \\ 
 3 & X : Season          && 51367 & 41 & 50450 & 39 & 50403 & 42 & 50395 & 46 \\ 
 4 & X * Season          && 51331 & 44 & 50418 & 42 & 50366 & 45 & 50360 & 49 \\ 
 5 & X + Milano           && 51473 & 15 & 50516 & 13 & 50492 & 16 & 50484 & 20 \\ 
 6 & X + Season + Milano  && 51442 & 18 & 50495 & 16 & 50468 & 19 & 50448 & 23 \\ 
 7 &(X : Season): Milano  && 50972 & 77 & 50064 & 75 & 50028 & 78 & 50015 & 82 \\ 
 8 & X : Season  * Milano && 50973 & 78 & 50065 & 76 & 50029 & 79 & 50016 & 83 \\ 
 9 &(X * Season) : Milano && 50945 & 83 & 50042 & 81 & 50002 & 84 & 49990 & 88 \\ 
10 &(X * Season)* Milano  && 50947 & 84 & 50043 & 82 & 50002 & 85 & 49991 & 89 \\ 
   \hline
\end{tabular}
\caption{Model selection by AIC. Df columns report model degrees of freedom. The minimum is attained at Model (9,4)}
\label{tab:AIC}
\end{table}

\begin{table}[ht]
\centering
\begin{tabular}{rllcccccccc}
& && \multicolumn{8}{c}{Skedastic model} \\
\cline{4-11}
& && \multicolumn{2}{c}{1} 
  & \multicolumn{2}{c}{2} 
  & \multicolumn{2}{c}{3}
  & \multicolumn{2}{c}{4}\\
\multicolumn{2}{c}{Regression model} && BIC & df & BIC & df & BIC & df & BIC & df \\ 
  \hline
 1 & X                   && 51757 & 14 & 50794 & 12 & 50788 & 15 & 50809 & 19 \\ 
 2 & X + Season          && 51714 & 17 & 50768 & 15 & 50756 & 18 & 50761 & 22 \\ 
 3 & X : Season          && 51652 & 41 & 50721 & 39 & 50696 & 42 & 50715 & 46 \\ 
 4 & X * Season          && 51637 & 44 & 50710 & 42 & 50679 & 45 & 50701 & 49 \\ 
 5 & X + Milano           && 51577 & 15 & 50607 & 13 & 50603 & 16 & 50623 & 20 \\ 
 6 & X + Season + Milano  && 51568 & 18 & 50606 & 16 & 50600 & 19 & 50608 & 23 \\ 
 7 &(X : Season): Milano  && 51507 & 77 & 50585 & 75 & 50571 & 78 & 50585 & 82 \\ 
 8 & X : Season  * Milano && 51515 & 78 & 50593 & 76 & 50579 & 79 & 50593 & 83 \\ 
 9 &(X * Season) : Milano && 51523 & 83 & 50606 & 81 & 50586 & 84 & 50602 & 88 \\ 
10 &(X * Season)* Milano  && 51531 & 84 & 50614 & 82 & 50594 & 85 & 50610 & 89 \\ 
   \hline
\end{tabular}
\caption{Model selection by BIC. Df columns report model degrees of freedom. The minimum is attained at Model (7,3).}
\label{tab:BIC}
\end{table}

AIC criterion selects Model (9,4) and BIC selects Model (7,3). Both show the importance of introducing multiplicative seasonality and Milano dummy. 
The difference between the AIC and BIC optima deserves some comments.
Considering the regression component, Model 7 provides different X-coefficients for each season and Milano. Model 9 also adds two additive terms for the season: one for LPV and one for Milano. Considering the skedastic component, it must be noted that the power coefficients $\delta_\mathcal{S}$ are very close to one. Hence, invoking the principle of Occam's Razor, we select the best BIC model - namely Model (7,3) - which has a determination coefficient $R^2 =0.764$.

Table \ref{tab:spat.corr} shows the residual correlation among the five stations. These correlations are smaller than the corresponding ones computed directly on PM2.5 observations, which are prone to spuriousness due to seasonality and meteorology.
Nevertheless, residual correlations are significantly different from zero at standard significance levels. 
In particular, Milano and Sannazzaro have the highest correlations, while Pavia has the smallest ones. From this perspective, we can say that Milano and Sannazzaro are more ``central'' in our data set, whilst Pavia is more ``peripheral''.

Considering the error variance structure, the estimated standard error is $\hat{\sigma}=0.05$, and the skedastic Model 3 has $\gamma$ estimates given by 12.3, 13.1, 9.3, and 12.1 for winter, spring, summer, and fall, respectively.

\subsection{Estimation results}

The grouped ANOVA in Table \ref{tab:anova} shows that all variables and factors of Model (7,3) are significant, confirming the validity of the BIC choice.
Table \ref{tab:beta.tTable} reports all estimated $\alpha$ and $\beta$ coefficients but the intercept, which is omitted for simplicity.
The first-order autoregressive component is always significantly positive and smaller in summer. In Milano, it is also small in winter.

For LPV stations, the NH3 coefficient is positive and highly significant for all seasons except in summer when NOX levels are high, which is quite a rare event.
In particular, it is larger in winter and spring, with the maximum in winter when NOX is high, consistent with the chemical interaction of the two species \citep{Veratti2023}.
Instead, the minimum is in summer when PM2.5 is lower.

At the Milano station, NH3 is generally non-significant, meaning that the low NH3 levels in the city have little influence on PM2.5. It is only significant in winter when NOX is high. In this case, it is also higher than the LPV value. This is also consistent with the mentioned interaction between NH3 and NOX \citep{Veratti2023}.

The NOX coefficient is always positive and highly significant both in and outside Milano. Interestingly, it is higher outside Milano.
The BLH coefficient is always significantly negative, but in spring and in Milano.
The wind always has a cleaning effect, with the WS coefficient being negative. It is highly significant for LPV stations.
We discuss these results further in Section \ref{sec:discussion}.

\subsubsection*{Meteorological confounders}
The role of temperature is interesting, as its coefficient is positive in summer and fall but negative in winter and spring. This was partly expected from the preliminary analysis shown in the first panel of Figure \ref{fig:SeasonalPM25vsALL}. In fact, the autumn behaviour is different when the other variables in the model are taken into account.

Rain has a cleansing effect in all seasons, but with markedly different coefficients. It is non-significant in summer and - in Milano - also in spring.
Lastly, RH has a significantly positive coefficient in all seasons except summer.

In conclusion, in LPV, almost all coefficients are significantly different from zero. In Milano, the relation between PM2.5 and NH3 is important only in winter.

\begin{table}[ht]
\centering
\begin{tabular}{lrrr}
  \hline
 & DF & F-stat & p-value \\ 
  \hline
  lag.PM25:Season:Milano   &   8 & 992.17 & 0.000 \\ 
  NH3.Low:Season:Milano    &   8 & 10.90 & 0.000 \\ 
  NH3.High:Season:Milano   &   8 & 62.99 & 0.000 \\ 
  NOX:Season:Milano 		  &   8 & 164.13 & 0.000 \\ 
  BLH:Season:Milano 		  &   8 & 21.76 & 0.000 \\ 
  WS:Season:Milano         &   8 & 21.45 & 0.000 \\ 
  T:Season:Milano          &   8 & 4.50 & 0.000 \\ 
  Rainy.Day:Season:Milano  &   8 & 7.67 & 0.000 \\ 
  RH:Season:Milano         &   8 & 20.58 & 0.000 \\ 
   \hline
\end{tabular}
\caption{Grouped ANOVA table of model in Equation \ref{eq:Model.Wilk}. The DF column reports the degrees of freedom.}
\label{tab:anova}
\end{table}

\begin{table}[!ht]
    \centering
    \begin{tabular}{ll cc c cc}
    \hline
    && \multicolumn{2}{c}{LPV} && \multicolumn{2}{c}{Milano} \\
    \cline{3-4}
    \cline{6-7}
        ~ & ~ & estimate & p-value &~& estimate & p-value \\ \hline
        lag(PM2.5) & Spring & 0.414 & 0.000 & ~ & 0.511 & 0.000 \\ 
        ~ & Summer & 0.339 & 0.000 &~& 0.264 & 0.000 \\ 
        ~ & Fall & 0.474 & 0.000 &~& 0.463 & 0.000 \\ 
        ~ & Winter & 0.415 & 0.000 &~& 0.310 & 0.000 \\ \hline
        NH3 - Low NOX & Spring & 0.273 & 0.000 &~& 0.176 & 0.014 \\ 
        ~ & Summer & 0.139 & 0.000 &~& 0.081 & ~ \\ 
        ~ & Fall & 0.240 & 0.000 &~& -0.088 & ~ \\ 
        ~ & Winter & 0.328 & 0.000 &~& 0.359 & ~ \\ \hline
        NH3 - High NOX & Spring & 0.181 & 0.003 &~& 0.145 & ~ \\ 
        ~ & Summer & 0.011 &   &~& -0.184 & ~ \\ 
        ~ & Fall & 0.240 & 0.000 &~& 0.086 & ~ \\ 
        ~ & Winter & 0.382 & 0.000 &~& 0.544 & 0.000 \\ \hline
        NOX & Spring & 0.186 & 0.000 &~& 0.100 & 0.000 \\ 
        ~ & Summer & 0.139 & 0.000 &~& 0.115 & 0.000 \\ 
        ~ & Fall & 0.112 & 0.000 &~& 0.084 & 0.000 \\ 
        ~ & Winter & 0.104 & 0.000 &~& 0.072 & 0.000 \\ \hline
        BLH & Spring & 0.000 &   &~& 0.000 & ~ \\ 
        ~ & Summer & -0.002 & 0.000 &~& -0.001 & ~ \\ 
        ~ & Fall & -0.003 & 0.000 &~& -0.002 & ~ \\ 
        ~ & Winter & -0.001 & 0.038 &~& 0.000 & ~ \\ \hline
        Wind Speed & Spring & -0.925 & 0.000 &~& -0.340 & ~ \\ 
        ~ & Summer & -1.106 & 0.000 &~& -0.506 & ~ \\ 
        ~ & Fall & -0.921 & 0.000 &~& -1.027 & 0.027 \\ 
        ~ & Winter & -1.372 & 0.000 &~& -2.190 & 0.003 \\ \hline
        Temperature & Spring & -0.143 & 0.001 &~& -0.020 & ~ \\ 
        ~ & Summer & 0.352 & 0.000 &~& 0.450 & 0.000 \\ 
        ~ & Fall & 0.156 & 0.002 &~& 0.230 & 0.011 \\ 
        ~ & Winter & -0.247 & 0.016 &~& -0.379 & 0.043 \\ \hline
        Rainy Day & Spring & -2.058 & 0.000 &~& -0.936 & ~ \\ 
        ~ & Summer & -0.231 &  &~& -0.733 & ~ \\ 
        ~ & Fall & -3.768 & 0.000 &~& -2.637 & 0.015 \\ 
        ~ & Winter & -4.925 & 0.000 &~& -6.856 & 0.000 \\ \hline
        Humidity & Spring & 0.094 & 0.000 &~& 0.040 & 0.047 \\ 
        ~ & Summer & 0.004 &  &~& -0.039 & ~ \\ 
        ~ & Fall & 0.080 & 0.000 &~& 0.060 & 0.006 \\ 
        ~ & Winter & 0.170 & 0.000 &~& 0.172 & 0.000 \\ \hline
    \end{tabular}
\caption{Estimated seasonal $\alpha$ and $\beta$ coefficients with their p-values in Lower Po Valley	 stations and Milano station. P-value columns: 0.000 denotes a p-value smaller than 0.0005; p-value is omitted when larger than 0.05 for improving readability.}
\label{tab:beta.tTable}
\end{table}

\begin{table}[ht]
\centering
\begin{tabular}{lccccc}
  \hline
 & Cremona & Milano & Sannazzaro & Schivenoglia & Pavia \\ 
  \hline
  Cremona       & 1.000 & 0.435 & 0.395 & 0.350 & 0.321 \\ 
  Milano        & 0.435 & 1.000 & 0.464 & 0.421 & 0.236 \\ 
  Sannazzaro    & 0.395 & 0.464 & 1.000 & 0.427 & 0.320 \\ 
  Schivenoglia  & 0.350 & 0.421 & 0.427 & 1.000 & 0.276 \\ 
  Pavia & 0.321 & 0.236 & 0.320 & 0.276 & 1.000 \\ 
   \hline
\end{tabular}
\caption{Residual correlation among stations.}
\label{tab:spat.corr}
\end{table}

\section{Discussion}\label{sec:discussion}
Due to the first-order autoregressive component in the model of Table \ref{tab:beta.tTable}, we interpret the NH3 (NOX) coefficients as the short-term sensitivity of PM2.5 to NH3 (NOX) after accounting for persistence, season, and meteorology.
The relationship between PM2.5 and precursors NH3 and NOX has several nonlinear components highlighted by our detailed local linear model. 

\subsubsection*{SQ1 - Is PM2.5 sensitive to NH3 and NOX?}
The answer is generally yes, but Milano station is different from LPV stations.

In particular, we found a highly significant NOX sensitivity for all stations and seasons. 
This is consistent with the fact that NOX sources are active in both urban and rural areas.
Notably, the sensitivity is higher in LPV. This may be related to two features. 
First, the PM2.5 persistence ($\alpha$) is larger in LPV than Milano, always but in spring.
Second, as shown by Table \ref{tab:elenco_stazioni}, the NOX level is higher in Milano and lower in LPV, while the opposite is true for PM2.5.

In addition, the NH3 sensitivity is generally highly significant in LPV, close to the farms. In Milano, where there are no agricultural NH3 sources, ammonia sensitivity is generally absent. An important exception is in winter when NOX is high. This condition is usually associated with a low boundary layer, reduced vertical air circulation, and possible horizontal air circulation in the lower atmosphere,
with transport and accumulation of pollutants.

\subsubsection*{SQ2 - Does this relation depend on the season?}
First, we observe that NH3 concentration has no large seasonal variation (Figure \ref{fig:AQ_timeseries}), contradicting the assumption that it depends strongly on the manure spreading calendar -  at least for the stations considered.

In LPV, we have already observed that winter has a slightly lower NH3 average, and the opposite is true for the Milano station. Nevertheless, PM2.5 is more sensitive to NH3 in winter in both areas, and for both high and low NOX levels.

\subsubsection*{SQ3 - Does the sensitivity to NH3 depend on NOX level?}
The effect of NOX level on NH3 sensitivity is complex. In winter, the sensitivity is higher when the NOX level is high, both in LPV and Milano.
In spring, the opposite is true, with higher NH3 sensitivity when NOX is low. It should be noted that high NOX in spring is a rare event, and there could be some confounding in action.

These findings are consistent with \cite{Veratti2023}:
\textit{``the analysis ... revealed that during the cold season, the efficiency of PM2.5 abatement tends to increase by increasing the emission reductions, while during summertime,
the same efficiency remains almost constant, or slightly decreases towards higher reduction strengths."}

\subsubsection*{SQ4 - Is this relation constant around Lombardy?}
In LPV stations, PM2.5 generally shows high significant sensitivity to NH3 when NOX is either low or high. 
Instead, at Milano station, PM2.5 shows little sensitivity to NH3, with significant positive coefficients only in spring when NOX is low and in winter when NOX is high. 
This is consistent with the fact that this station is located in a large metropolitan area, far from farms. 
Even though it is an urban background (UB) station, PM2.5 is sensitive to NOX due to car traffic and house heating.
It is interesting to note that the maximum NOX sensitivity is in summer.

So, even if we cannot fully answer SQ4 because of the limited network, we can identify two patterns (Milano and LPV) that differ geographically and in terms of sensitivity.

\section{Conclusions}\label{sec:conclusions}
This observational study is based on available data on concentrations of PM2.5, NH3, and NOX in Lombardy - namely five stations.
Using a statistical model that takes into account heteroskedasticity and spatial correlation, we estimated the sensitivity of PM2.5 to NOX and NH3.
Our results are consistent with those obtained using ammonia emissions \citep{Thunis2021,Veratti2023,otto2023spatiotemporal}.

The sensitivity of PM2.5 results is highly seasonal and differs between Milano and the Lower Po Valley stations. 
An air quality plan to reduce pollution should consider the following:
\begin{itemize}
\item Since NH3 sensitivity depends on NOX levels, a joint reduction policy would provide the best results.
\item At least for the five stations considered, there is limited evidence of NH3 seasonality. Therefore, an emission reduction policy focused only on manure spread practices is inadequate. Indeed, it should focus on the entire manure cycle.
\item The reduction focus should be in winter when PM2.5 concentration and NH3 sensitivity are higher.
\end{itemize}

\begin{acknowledgement}
This research was co-funded by Fondazione Cariplo under the grant 2020–4066 \textit{“AgrImOnIA: the
impact of agriculture on air quality and the COVID-19 pandemic”} from the \textit{“Data Science for
Science and Society”} program and by the European Union -  NextGenerationEU, in the framework of the \textit{“GRINS - Growing Resilient, INclusive and Sustainable”} project (GRINS PE00000018 – CUP F83C22001720001). The views and opinions expressed are solely those of the authors and do not necessarily reflect those of the European Union, nor can the European Union be held responsible for them.
\end{acknowledgement}

\bibliography{bibliography}   	

\end{document}